\documentclass[nofootinbib, showpacs, preprintnumbers, amsmath,amssymb, floatfix, reprint,superscriptaddress, prl, aps, nobibnotes]{revtex4-1}
\usepackage{hyperref, graphicx}
\usepackage{dcolumn}
\usepackage{bm}
\usepackage{siunitx}
\usepackage{textcomp, gensymb}
\usepackage{parskip} 
\newcommand{\posem}[0]{$\beta^+$}
\begin{document}

\preprint{APS/123-QED}

\title{\textbf{Precision measurement of positron decay modes of $^{125}\text{Xe}$ \\ in the LUX-ZEPLIN experiment} 
}%



\author{D.S.~Akerib}
\affiliation{SLAC National Accelerator Laboratory, Menlo Park, CA 94025-7015, USA}
\affiliation{Kavli Institute for Particle Astrophysics and Cosmology, Stanford University, Stanford, CA  94305-4085 USA}

\author{A.K.~Al Musalhi}
\affiliation{University College London (UCL), Department of Physics and Astronomy, London WC1E 6BT, UK}

\author{F.~Alder}
\affiliation{University College London (UCL), Department of Physics and Astronomy, London WC1E 6BT, UK}

\author{B.J.~Almquist}
\affiliation{Brown University, Department of Physics, Providence, RI 02912-9037, USA}

\author{C.S.~Amarasinghe}
\affiliation{University of California, Santa Barbara, Department of Physics, Santa Barbara, CA 93106-9530, USA}

\author{A.~Ames}
\affiliation{SLAC National Accelerator Laboratory, Menlo Park, CA 94025-7015, USA}
\affiliation{Kavli Institute for Particle Astrophysics and Cosmology, Stanford University, Stanford, CA  94305-4085 USA}

\author{T.J.~Anderson}
\affiliation{SLAC National Accelerator Laboratory, Menlo Park, CA 94025-7015, USA}
\affiliation{Kavli Institute for Particle Astrophysics and Cosmology, Stanford University, Stanford, CA  94305-4085 USA}

\author{N.~Angelides}
\affiliation{University of Zurich, Department of Physics, 8057 Zurich, Switzerland}

\author{H.M.~Ara\'{u}jo}
\affiliation{Imperial College London, Physics Department, Blackett Laboratory, London SW7 2AZ, UK}
\affiliation{STFC Rutherford Appleton Laboratory (RAL), Didcot, OX11 0QX, UK}

\author{J.E.~Armstrong}
\affiliation{University of Maryland, Department of Physics, College Park, MD 20742-4111, USA}

% ...

\collaboration{The LZ Collaboration}


\begin{abstract}
The radioisotope $^{125}\text{Xe}$ is a short-lived ($T_{1/2}\sim16.9~h$) activation product of the neutron calibrations performed in the LUX-ZEPLIN experiment. Subsequently, $^{125}$Xe decays primarily ($>99\%$) via electron capture, but positron emission has been confirmed by direct measurement to at least the 243~keV level of $^{125}\text{I}$. An additional decay to the 188~keV level is expected from triple-coincident measurements of the annihilation and relaxation $\gamma$ rays, but has not been directly confirmed. 
By utilizing multiple-scatter event analysis and the pre-activation data to constrain backgrounds, this work reports positron emission with a statistical significance of 5.5~$\sigma$. This corresponds to a total branching ratio of $0.29\pm0.08_{~\text{stat.}}\pm0.04_{~\text{sys.}}$~\%, and is the first constraint to the individual branching levels of $^{125}\text{I}$.
\end{abstract}

\date{\today}

\maketitle



\begin{figure}[!htp]
    \centering
    \includegraphics[width=0.65\columnwidth]{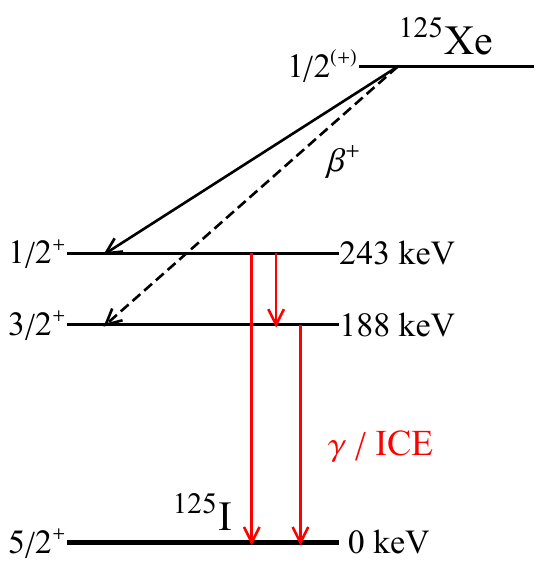}
    \caption{A simplified \posem-only decay level scheme of $^{125}\text{Xe}$ is shown with excited energies in reference to the $^{125}\text{I}$ ground state. The decay to the 243 keV level (solid black) has been directly confirmed, while the decay to the 188 keV level (dashed black) has only been inferred from a coincidence measurement of the 188 keV and annihilation $\gamma$ rays. This would also arise from the intermediate $243~\text{keV}\rightarrow188~\text{keV}$ de-excitation, which may be via $\gamma$-emission or electron internal conversion (red lines).}
    \label{fig:125i}
\end{figure}

The LUX-ZEPLIN (LZ) experiment is a direct-detection dark matter search employing a dual-phase xenon time projection chamber (TPC), optimized for direct searches of weakly interacting massive particles (WIMPs)~\cite{LZ:2019sgr}. LZ is acquiring data as of late 2021, achieving the world-leading sensitivity to WIMP-nucleon interactions via keV-scale nuclear recoils (NRs) with a 4.2 tonne-year exposure~\cite{LZ:2024zvo}. For the electron recoil (ER) channel, LZ is sensitive to a wide range of possible beyond-Standard Model processes with energy depositions between several keV to several MeV~\cite{LZ:2019qdm, LZ:2021blo,  LZ:2023poo}. 
\\ \indent
Detector response tuning of scintillation yields and extraction efficiencies utilizes multiple calibration sources described extensively in previous publications~\cite{Mount:2017qzi, LZ:2024bsz}. NR calibrations, such as the deterium-deterium (DD) fusion source used in this analysis, irradiate the target liquid xenon (LXe), during which neutrons may thermalize or capture in-flight. This results in a period of elevated decay rates of unstable xenon isotopes after the calibration. 
\\ \indent
The high neutron capture cross-section of $^{124}\text{Xe}$ relative to other stable isotopes results in a large production of the short-lived $^{125}\text{Xe}$ despite its low natural abundance~\cite{Shustov:2016hum}. $^{125}\text{Xe}$ primarily decays via electron-capture (EC) with a half-life of about 16.9~hours~\cite{ensdf-125}. Positron emission (\posem~decay) of $^{125}\text{Xe}$ has been directly confirmed to the 243~keV excited state of $^{125}\text{I}$~\cite{Ludziejewski:1969}, with the decay scheme illustrated in Fig.~\ref{fig:125i}. This analysis observes the combined signature of several energy depositions: the $\beta^+$ kinetic energy, de-excitation $\gamma$-rays or their respective internal conversion electrons, and a pair of $e^+e^-$ annihilation $\gamma$ rays. The following sections will describe the adaptations of the event reconstruction and selection for multiple-scatter events, the expected signal and background sources, and the analysis of post-calibration events.

\begin{figure}[!h]
    \centering
    \includegraphics[width=\columnwidth]{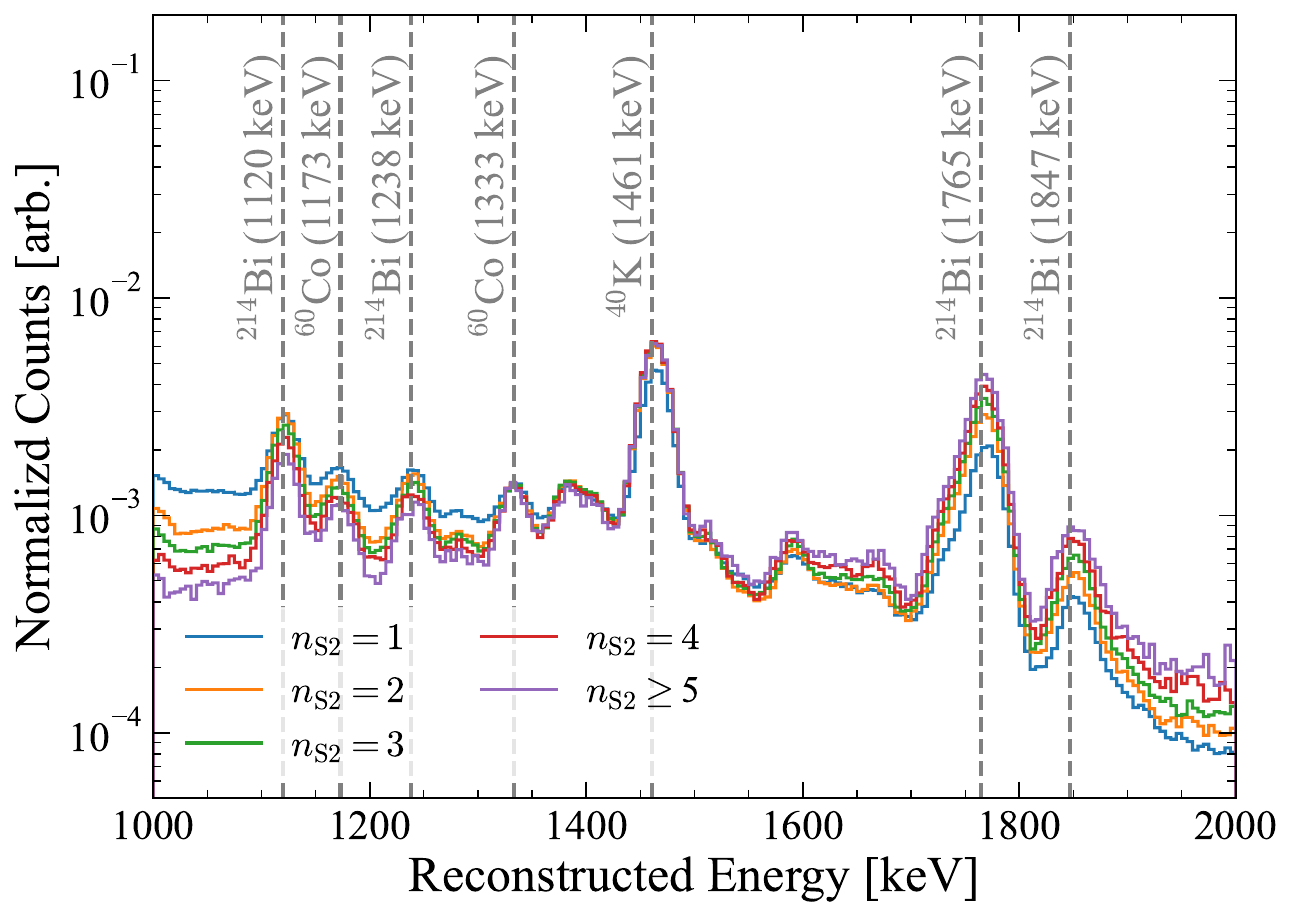}

    \caption{Reconstructed energy for WS2024 background data, showing consistency in reconstructed energy for single-site and multiple-site high-energy events across S2 multiplicity. The energies of several known monoenergetic gamma-ray emission lines are indicated (dashed gray).}
    \label{fig:rec_en}
\end{figure}

A complete description of the LZ experiment can be found in existing publications~\cite{Mount:2017qzi, LZ:2019sgr, LZ:2022ysc, LZ:2022lsv, LZ:2025xxf}. In the LZ detector, particles interacting with the LXe target deposit energy as scintillation light, ionization, and unobserved heat. The prompt scintillation light (S1) is recorded on top and bottom arrays of photomultiplier tubes (PMTs). A nearly uniform electric field drifts the ionization electrons upwards to the gaseous xenon region, where a stronger field extracts them from the surface and generates delayed electroluminescence (S2) via proportional scintillation. The extracted signals, reported in units of photons detected (phd)~\cite{Faham:2015kqa}, provide reconstructed energy, position, and interaction type discrimination. 
\\ \indent Depth information ($z$) is reconstructed from the electron drift-time ($\tau$), and radial information from the PMT hit pattern of the S2 ($x, y$). Multiple simultaneous relativistic particles, e.g. X-ray and $\gamma$ ray from a decay, or a single relativistic particle scattering multiple times, result in merged S1, while the S2 pulses may be separately resolvable, provided separation of a few mm in $z$. 
This provides the individual S2 position and energy reconstruction, which is utilized in this search.
\\ \indent
The standard LZ event reconstruction procedure applies corrections for detector effects and detection efficiencies, assuming a single-site interaction. We modify the corrected variables to consider the separate high-energy depositions in the detector. To avoid PMT saturation from the proportional scintillation at higher energies, only the pulse area of the bottom array ($\text{S2}^{bot}$) is used for energy reconstruction. Each S2 pulse is then corrected with the standard LZ spatial detector response corrections to yield  $\text{S2}c^{bot}_{i}$. All S2 will have their own $\tau$-dependent correction factor, resulting in multiple corrected $\text{S}1c_i$. Total corrected S1 area ($\text{S1}c$) and centroid position variables ($x,~y,~z$) are computed using an \mbox{$\text{S2}c_{i}^{bot}$-weighted} sum, 
\begin{equation}
\chi= \frac{\sum_{i} \chi_i \cdot \text{S2}c^{bot}_i}{\sum_{i} \text{S2}c^{bot}_i}.
\end{equation}
From the LXe quanta work function at detector conditions, $W = 13.5\textrm{ quanta}/\textrm{eV}$~\cite{Szydagis:2021hfh}, the total event energy is reconstructed as
\begin{equation}
E = W \cdot \left(\frac{\text{S1}c}{g_1} + \frac{\sum_i \text{S2}c^{bot}_i} {g_{2}^{bot}}\right).
\end{equation}
The scintillation and ionization gain factors were determined through the Doke plot method \cite{Aprile:2009dv, Baudis:2021dsq} to be $g_1 = 0.1132\textrm{ phd}/\textrm{emitted~photon}$ and $g_2^{bot}  =11.55\textrm{ phd}/\textrm{ionization electron}$, respectively. The discrepancy of these factors from existing publications \cite{LZ:2024zvo} is attributed to variation in multiple-scatter interactions reconstruction and is considered as a systematic effect on the final analysis result. The agreement of reconstructed energy with these factors for various background $\gamma$ peaks with different numbers of resolved sites ($n_{\text{S2}}$) is shown in Fig.~\ref{fig:rec_en}. 

The $^{125}\text{Xe}$ activated dataset consists of a cumulative 1.5 live-day acquisition immediately following a continuous 11-day exposure of deuterium-deuterium neutrons reflected off an aluminum cell filled with deuterated benzene liquid scintillator. The reflector configuration provides a peaked incident kinetic energy of $349\pm3~\text{keV}$ with a full-width at half-maximum of $79\pm2~\text{keV}$, further discussed in Ref.~\cite{LZ:2024bsz}. During a $^{125}\text{Xe}$ $\beta^+$ decay, the creation of the positron and the loss of an orbital electron results in reduced decay energy available for the positron and excited $^{125}\text{I}$: $\tilde{Q}\approx Q - 1022~\text{keV} \approx 623~\text{keV}$. Upon annihilation of the $\beta^+$, a pair of 511 keV $\gamma$ rays is created, which are likely to additionally Compton scatter before photoelectric absorption. Assuming the decay products are fully contained in the instrumented volume, the energy of this signal ranges from about 1210~keV to 1645~keV. For an additional constraint on the EC branches, a widened range of 1000~keV to 1650~keV is considered as the energy window.
\\ \indent
The predominant pre-existing backgrounds before neutron activation are high-energy $\gamma$ rays undergoing multiple Compton scatters, mainly 1173~keV and 1333~keV lines from  $^{60}$Co, and a 1461~keV line from  $^{40}$K over a continuous background Compton spectrum. A $^{125}\text{Xe}$-free background model was constructed using an 85.2 cumulative live-day subset of WIMP search data in the WS2024~\cite{LZ:2024zvo} detector conditions before any neutron calibrations.
\\ \indent
Post-activation, the dominant decay mode of EC includes a few prominent transitions that reside in the energy window, as shown in Table~\ref{tab:table1}. Upon filling of the orbital shell vacated by the EC, an X-ray or Auger electron of corresponding energy is emitted, typically from the K or L shells, adding to the total energy of these transitions. Additionally, the activation products $^{135}\text{Xe}$ and $^{137}\text{Xe}$ reside in energy window, with their respective half-lives of 9.14~h and 3~min. $^{135}\text{Xe}$ has a negligible contribution to excited states that reside in the energy window. This has been evaluated at less than one $^{135}\text{Xe}$ decay in the final analysis. In contrast, the wide energy spectrum of $^{137}\text{Xe}$ $\beta^-$ decay will arise in the energy window. However, the primary neutron activation of xenon occurs in the outer layers, and the slow radial circulation with respect to the short half-life prevents a significant activation in the inner fiducial volume~\cite{LZ:2025xxf}.
\begin{table}[!htbp]
\caption{\label{tab:table1}
Prominent $^{125}\text{Xe}$ EC decay modes coincident in the energy window.}
\begin{ruledtabular}
\begin{tabular}{cccccccc}
Energy Level [keV]\footnotemark[1] & Branching Ratio [\%]\footnotemark[2]\\
\hline
\noalign{\vskip 1mm}    
1007.5& 0.34\\
1082.8& 0.02 \\
1089.9& 1.89 \\
1180.9& 1.00 \\
1264.0& 0.09 \\
1381.6& 0.37 \\
1442.8& 0.03 \\
1562.4& 0.005 \\
\end{tabular}
\end{ruledtabular}
\footnotetext[1]{Energy of X-ray/Auger electron from filling of capture vacancy not included.}
\footnotetext[2]{Branching ratios adopted from Ref.~\cite{ensdf-125}.
}

\end{table}




\begin{figure*}[!t]
    \centering
    \includegraphics[width=1.8\columnwidth]{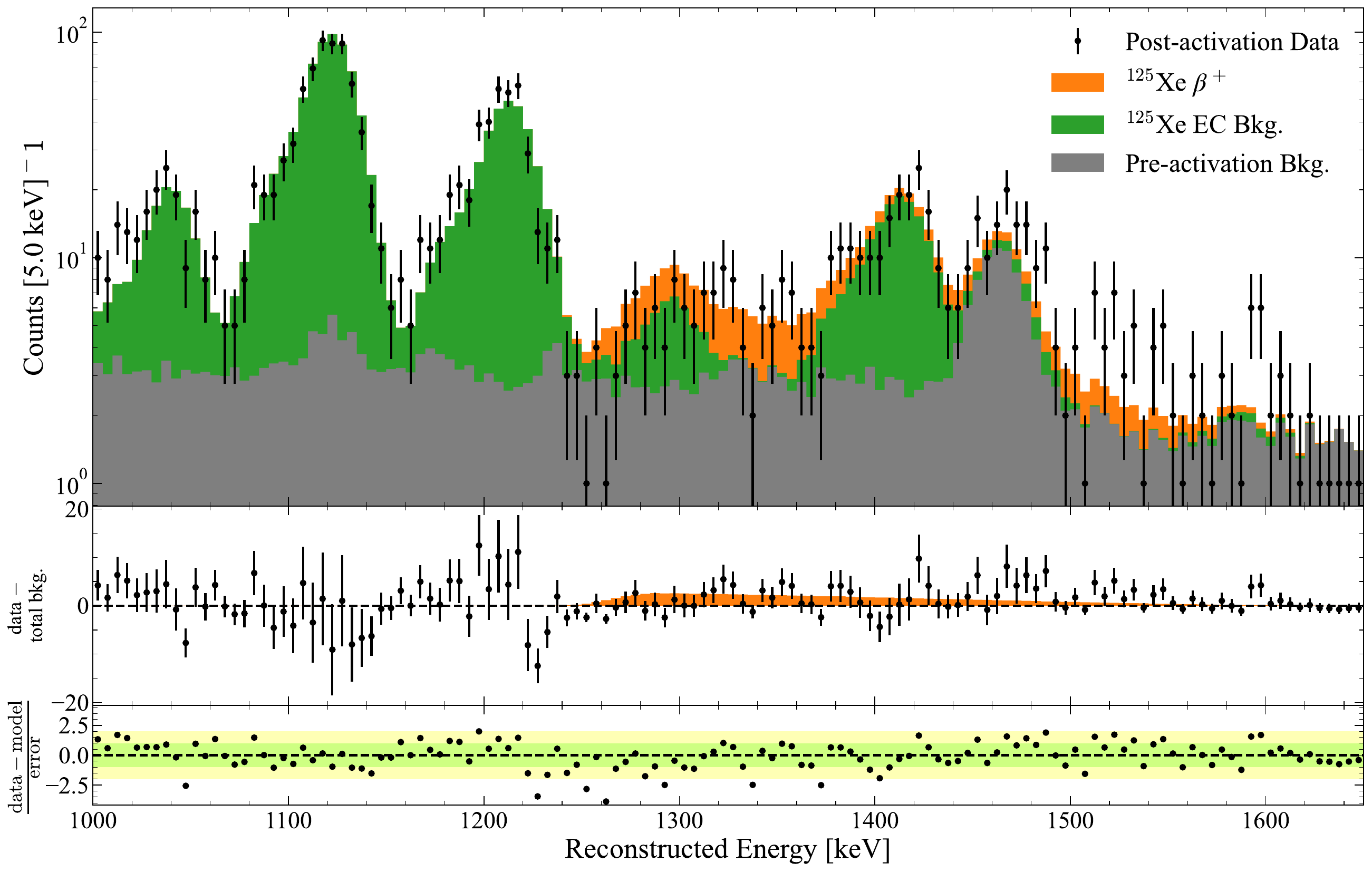}

    \caption{Resulting best-fit from the likelihood analysis shown in the top panel, with two additional residual panels shown below. The top panel shows the resulting post-activation data points (black), and the pre-activation background (grey), $^{125}\text{Xe EC}$ (green), and $^{125}\text{Xe }\beta^{+}$ (orange) fit components. The middle residual plot shows activated data points subtracted by the total background fit (black), overlaid with the individual $^{125}\text{Xe }\beta^{+}$ fit (orange). The bottom residual panel shows the post-activation data points subtracted by the total over their Poisson error, showing the agreement of the best fit across the energy spectrum for the majority of points within one (green-band) or two deviations (yellow-band).}
    \label{fig:fit}
\end{figure*}
The expected $\beta^+$ decay signal is dominated by interactions with multiple resolved scatters; however including single resolved scatters allows for a better constraint on the EC activation background and also remaining robust against inaccuracies in the modeling of single and multiple scatter ratios.
To remove spurious noise and pulses we enforce an event requirement for all interaction sites of \mbox{$\text{S2}_i>645~\text{phd}$~(14.5 extracted electrons)}, and total event requirement of \mbox{$3500~\text{phd}<\text{S1}c < 16000~\text{phd}$} and \mbox{$10^5~\text{phd} <\text{S2}c^{bot} <3.16\times10^{6}~\text{phd}$}. To ensure valid reconstruction of all S2s a physical radius and depth cut is applied, \mbox{$R_i <72.8~\text{cm}$} and \mbox{$0~
\text{cm} < z_i < 145.6~\text{cm}$}. A fiducial volume of about one-tonne is selected as defined in previous background studies \cite{LZ:2022ysc}, using the weighted centroid position variables with \mbox{$r< 45~\text{cm}$} and \mbox{$35~\text{cm} < z < 95~\text{cm}$}.  The fiducial cut further reduces the measured external $\gamma$-induced backgrounds in the final analysis by 99.6\% to a rate of about $3~\text{mHz}$.

Events passing the analysis selections are included in a reconstructed energy-binned likelihood with simultaneous maximization of two terms: one for the pre-activation period and one for the post-activation period. The post-activation term contains simulated background and signal components assuming a uniform spatial distribution, using the LZ \textsc{geant4}-based package \textsc{baccarat} \cite{GEANT4:2002zbu, LZ:2020zog} and processed through a detector response model utilizing \textsc{nest} parameters tuned on high-energy background sources\footnotemark[1]~\cite{Szydagis:2011tk}. As the EC branching modes have been measured with higher precision \cite{ensdf-125}, the $\beta^{+}$ branching modes can be freely floated in the fit and subsequently provides a constraint with the post-fit value. To allow the pre-activation dataset to constrain the non-activation backgrounds, each reconstruction energy-bin background expectation is profiled analytically, with an approach similar to Ref.~\cite{Barlow:1993dm}. Discovery significance and confidence intervals are computed assuming an asymptotic log-likelihood ratio test statistic \cite{Cowan:2010js}. Statistical uncertainties of signal and background parameters reported below are similarly computed with log-likelihood profiling.
\\ \indent
The post-activation data, along with the fitted model of the measurement, are shown in the top panel of Fig.~\ref{fig:fit}. The residuals, in the middle panel, show no significant deviations, and the goodness-of-fit using a binned Poisson likelihood $\chi^2$ test is acceptable ($p=0.057$). Best-fit parameters, including individual branching ratio constraints of the 188~keV and 243~keV levels, are shown in Table~\ref{tab:table4}. 
The predominant source of systematic errors on these measurements was identified as the existing constraints on the EC measured modes and the uncertainty on the $g_1$ and $g_{2}^{bot}$ efficiencies, and is included in the signal best-fit. 
The best-fit  $^{125}$Xe \posem~signal is  $96^{+29}_{-27}$~counts, corresponding to a branching ratio of \posem~of $0.29\pm0.08_{~\text{stat.}}\pm0.04_{~\text{sys.}}$~\%, providing increased precision of this process while consistent with existing literature value of $0.3\pm0.1_{~\text{tot.}}$~\% \cite{ensdf-125}. The discovery significance with respect to the EC-only null hypothesis is 5.5~$\sigma$.
\\ \indent
This is the first simultaneous measurement of all of the emitted quanta from the $^{125}\text{Xe }\beta^+$ decay, including the direct measurement of the kinetic energy deposited by the emitted $\beta^+$. Additionally, this measurement is the first constraint on the individual branching ratios of the 188~keV and 243~keV levels. This measurement demonstrates the power of LXe TPCs in understanding complex event topologies, and provides improved constraints on the total $^{125}\text{Xe}$ $\beta^+$ decay. LZ demonstrates the ability for in-situ production of $\beta^+$ decay via neutron activation on xenon, which may be used as a calibration source for improved understanding of the capability in future rare $\beta^+$ decay searches~\cite{Wittweg:2020fak}. 
\begin{table}[htpb]
\begin{ruledtabular}
\caption{\label{tab:table4}
Number of events from various sources and other fit results from the neutron-activated dataset likelihood.}
\begin{tabular}{cccccccc}
Source & Fit Results [counts]\\
\hline
\noalign{\vskip 1mm}    
Pre-activation Bkg. & $406^{+30}_{-30}$ \\
\noalign{\vskip 0.5mm}    

$^{125}$Xe EC Bkg. & $1229^{+55}_{-54}$\phantom{0} \\
\noalign{\vskip 0.5mm}    

$^{125}$Xe $\beta^+$ 188~keV & \phantom{00}$0^{+17}_{-0}$ \\
\noalign{\vskip 0.5mm}    

$^{125}$Xe $\beta^+$ 243~keV & \phantom{0}$96^{+29}_{-27}$ \\
\noalign{\vskip 1mm}    

\hline
\noalign{\vskip 1mm}    
Total & $1731^{+71}_{-67}$\phantom{0}\\
\noalign{\vskip 0.5mm}    
BR($\beta^+$): 188 keV & \phantom{00.}$0.00^{+0.05}_{-0.00}\%$\\
\noalign{\vskip 0.5mm}    
BR($\beta^+$): 243 keV & \phantom{00.}$0.29^{+0.09}_{-0.09}\%$\\
\noalign{\vskip 0.5mm}    
BR($\beta^+$): Total\phantom{ ke} & \phantom{.00}$0.29^{+0.09}_{-0.09}\%$\\
\noalign{\vskip 0.5mm}    
$\sigma_{sig}$ & 5.47\\
\noalign{\vskip 0.5mm}    
$p(\chi^2)$ & 0.057\\

\end{tabular}
\end{ruledtabular}

\end{table}

\footnotetext[1]{The \textsc{NEST} implementation assumes one yields model for all ER interactions, which is suitable for a combined energy-scale analysis.}


\begin{acknowledgments}

\emph{Acknowledgements--}
The research supporting this work took place in part at the Sanford Underground Research Facility (SURF) in Lead, South Dakota. Funding for this work is supported by the U.S. Department of Energy, Office of Science, Office of High Energy Physics under Contract Numbers DE-AC02-05CH11231, DE-SC0020216, DE-SC0012704, DE-SC0010010, DE-AC02-07CH11359, DE-SC0015910, DE-SC0014223, DE-SC0010813, DE-SC0009999, DE-NA0003180, DE-SC0011702, DE-SC0010072, DE-SC0006605, DE-SC0008475, DE-SC0019193, DE-FG02-10ER46709, UW PRJ82AJ, DE-SC0013542, DE-AC02-76SF00515, DE-SC0018982, DE-SC0019066, DE-SC0015535, DE-SC0019319, DE-SC0024225, DE-SC0024114, DE-AC52-07NA27344, \& DE-SC0012447. This research was also supported by U.S. National Science Foundation (NSF); the UKRI’s Science \& Technology Facilities Council under award numbers ST/W000490/1, ST/W000482/1, ST/W000636/1, ST/W000466/1, ST/W000628/1, ST/W000555/1, ST/W000547/1, ST/W00058X/1, ST/X508263/1, ST/V506862/1, ST/X508561/1, ST/V507040/1 , ST/W507787/1, ST/R003181/1, ST/R003181/2,  ST/W507957/1, ST/X005984/1, ST/X006050/1; Portuguese Foundation for Science and Technology (FCT) under award numbers PTDC/FIS-PAR/2831/2020; the Institute for Basic Science, Korea (budget number IBS-R016-D1); the Swiss National Science Foundation (SNSF) under award number 10001549. This research was supported by the Australian Government through the Australian Research Council Centre of Excellence for Dark Matter Particle Physics under award number CE200100008. We acknowledge additional support from the UK Science \& Technology Facilities Council (STFC) for PhD studentships and the STFC Boulby Underground Laboratory in the U.K., the GridPP \cite{GridPP:2006wnd, Britton:2009ser} and IRIS Collaborations, in particular at Imperial College London and additional support by the University College London (UCL) Cosmoparticle Initiative, and the University of Zurich. We acknowledge additional support from the Center for the Fundamental Physics of the Universe, Brown University. K.T. Lesko acknowledges the support of Brasenose College and Oxford University. The LZ Collaboration acknowledges the key contributions of Dr. Sidney Cahn, Yale University, in the production of calibration sources. This research used resources of the National Energy Research Scientific Computing Center, a DOE Office of Science User Facility supported by the Office of Science of the U.S. Department of Energy under Contract No. DE-AC02-05CH11231. We gratefully acknowledge support from GitLab through its GitLab for Education Program. The University of Edinburgh is a charitable body, registered in Scotland, with the registration number SC005336. The assistance of SURF and its personnel in providing physical access and general logistical and technical support is acknowledged. We acknowledge the South Dakota Governor's office, the South Dakota Community Foundation, the South Dakota State University Foundation, and the University of South Dakota Foundation for use of xenon. We also acknowledge the University of Alabama for providing xenon. For the purpose of open access, the authors have applied a Creative Commons Attribution (CC BY) license to any Author Accepted Manuscript version arising from this submission. Finally, we respectfully acknowledge that we are on the traditional land of Indigenous American peoples and honor their rich cultural heritage and enduring contributions. Their deep connection to this land and their resilience and wisdom continue to inspire and enrich our community. We commit to learning from and supporting their effort as original stewards of this land and to preserve their cultures and rights for a more inclusive and sustainable future. 

Disclaimer:
This report was prepared as an account of work sponsored by an agency of the United States Government. Neither the United States Government nor any agency thereof, nor any of their employees, makes any warranty, express or implied, or assumes any legal liability or responsibility for the accuracy, completeness, or usefulness of any information, apparatus, product, or process disclosed, or represents that its use would not infringe privately owned rights. Reference herein to any specific commercial product, process, or service by trade name, trademark, manufacturer, or otherwise does not necessarily constitute or imply its endorsement, recommendation, or favoring by the United States Government or any agency thereof. The views and opinions of authors expressed herein do not necessarily state or reflect those of the United States Government or any agency thereof.

\end{acknowledgments} 

\bibliography{default}

\end{document}